\documentclass[aps, prl, floatfix, longbibliography, nofootinbib, showpacs, twocolumn, superscriptaddress]{revtex4-1}

\usepackage{mathrsfs}
\usepackage{amsmath}
\usepackage{amssymb}
\usepackage{bbold}
\usepackage{color}
\usepackage{graphicx}
\usepackage{soul}

\usepackage[mathscr,scaled=1.15]{urwchancal}
\DeclareFontFamily{OT1}{pzc}{}
\DeclareFontShape{OT1}{pzc}{m}{it}%
{<-> s * [1.15] pzcmi7t}{}
\DeclareMathAlphabet{\mathpzc}{OT1}{pzc}{m}{it}

\definecolor{purple}{rgb}{0.5,0,0.5}
\definecolor{blue}{rgb}{0.0,0,0.9}

\begin{document}

\title{Partonic structure of neutral pseudoscalars via two photon transition form factors}

\author{Kh{\'e}pani Raya}
\email{khepani@ifm.umich.mx}
\affiliation{Instituto de F\'{\i}sica y Matem\'aticas, Universidad
Michoacana de San Nicol\'as de Hidalgo\\
Edificio C-3, Ciudad Universitaria, C.P. 58040,
Morelia, Michoac\'an, M{\'e}xico}

\author{Minghui Ding}
\email{mhding@nankai.edu.cn}
\affiliation{School of Physics, Nankai University, Tianjin 300071, China}

\author{Adnan Bashir}
\email{adnan@ifm.umich.mx}
\affiliation{Instituto de F\'{\i}sica y Matem\'aticas, Universidad
Michoacana de San Nicol\'as de Hidalgo\\
Edificio C-3, Ciudad Universitaria, C.P. 58040,
Morelia, Michoac\'an, M{\'e}xico}

\author{Lei Chang}
\email{leichang@nankai.edu.cn}
\affiliation{School of Physics, Nankai University, Tianjin 300071, China}

\author{Craig~D.~Roberts}
\email{cdroberts@anl.gov}
\affiliation{Physics Division, Argonne National Laboratory, Argonne
Illinois 60439, USA}

\date{19 October 2016}

\begin{abstract}
The $\gamma \gamma^\ast \to \eta_{c,b}$ transition form factors are computed using a continuum approach to the two valence-body bound-state problem in relativistic quantum field theory, and thereby unified with equivalent calculations of electromagnetic pion elastic and transition form factors.
The resulting $\gamma \gamma^\ast \to \eta_c$ form factor, $G_{\eta_c}(Q^2)$, is consistent with available data:  significantly, at accessible momentum transfers, $Q^2 G_{\eta_c}(Q^2)$ lies well below its conformal limit.  These observations confirm that the leading-twist parton distribution amplitudes (PDAs) of heavy-heavy bound-states are compressed relative to the conformal limit.
A clear understanding of the distribution of valence-quarks within mesons thus emerges; a picture which connects Goldstone modes, built from the lightest-quarks in Nature, with systems containing the heaviest valence-quarks that can now be studied experimentally, and highlights basic facts about manifestations of mass within the Standard Model.
\end{abstract}


\smallskip


\maketitle


\noindent\emph{I.$\;$Introduction}\,---
%
The properties of pseudoscalar mesons provide a unique window on the Standard Model.  For example, $\gamma \gamma^\ast \to \pi^0$, neutral pion production in two photon fusion \cite{Behrend:1990sr, Gronberg:1997fj, Aubert:2009mc, Uehara:2012ag},
ties physics associated with a nonperturbative anomaly \cite{Adler:1969gk, Bell:1969ts} to that connected with collinear factorisation in hard-scattering processes as demonstrated through the application of perturbative quantum chromodynamics (pQCD) \cite{Farrar:1979aw, Lepage:1979zb, Efremov:1979qk, Lepage:1980fj}.  Simultaneously, pion properties provide a clean probe of the mechanisms responsible for the generation of more than 98\% of visible mass in the universe \cite{Horn:2016rip, Roberts:2016vyn}; and yet, from another perspective, one might view the production of charm-anticharm systems via gluon-gluon fusion as yielding valuable, complementary information on this same subject \cite{Kharzeev:1995ij, Kharzeev:1998bz}.  Such processes are echoed in the reaction $\gamma \gamma^\ast \to \eta_c$: measured at photon virtualities in the range $2\lesssim Q^2 \lesssim 50$GeV$^2$ \cite{Lees:2010deBaBar}, a subject of phenomenological analyses \cite{Kroll:2010bf, Lucha:2012ar, Bedolla:2016yxq}, it is often supposed to yield information on the strong running-coupling at the charm-quark mass, which can be used to inform and refine effective field theories developed for application to systems involving heavy quarks \cite{Brambilla:2010cs, Bevan:2014iga}.

Measurements of pseudoscalar meson production via two-photon fusion are challenging.  They typically involve the study of $e^-$-$e^+$ collisions, in which one of the outgoing fermions is detected after a large-angle scattering whilst the other is scattered through a small angle and, hence, undetected.  The detected fermion is assumed to have emitted a highly-virtual photon, the undetected fermion, a soft-photon; and these photons are supposed to fuse and produce the final-state pseudoscalar meson.  There are many possible background processes and loss mechanisms in this passage of events, and thus ample room for systematic error, especially as $Q^2$ increases \cite{Bevan:2014iga}.  The potential for such errors plays a large part in the controversy surrounding the most recent measurements of $\gamma \gamma^\ast \to \pi^0$ \cite{Aubert:2009mc, Uehara:2012ag}, which exhibit incompatible trends in their evolution with photon virtuality \cite{Stefanis:2012yw}.  It does not, however, appear to play a significant role in the debate over whether effective field theory methods can be used to understand contemporary $\gamma \gamma^\ast \to \eta_c$ data \cite{Lees:2010deBaBar}: whilst leading-order (LO) and next-to-leading-order (NLO) analyses in non-relativistic QCD (nrQCD) seem adequate, next-to-next-to-leading-order (NNLO) corrections very seriously disrupt agreement with experiment \cite{Feng:2015uha}.  One should therefore also question related predictions for $\gamma \gamma^\ast \to \eta_b$.

A consolidated explanation of all three transition form factors within a single theoretical approach would facilitate a resolution of these disputes.  Herein, therefore, we employ a symmetry-preserving framework for the study of strong-interaction bound-states \cite{Cloet:2013jya, Roberts:2015lja} in an attempt to provide a unified description of the $\gamma \gamma^\ast \to \pi^0$ and $\gamma \gamma^\ast \to \eta_c$  transitions, along with a prediction for $\gamma \gamma^\ast \to \eta_b$.  In so doing, we will reveal how these form factors provide insights into the nature of momentum sharing between the  valence quanta in these bound-states.


\smallskip

\noindent\emph{II.$\;$Transition Form Factors: Formulation}\,---
The transition $\gamma\gamma^\ast  \to M_5$ is described by a single scalar function, required to fully express the amplitude:
\begin{align}
%
T_{\mu\nu}(k_1,k_2) & = \tfrac{e^2}{4\pi^2} \epsilon_{\mu\nu\alpha\beta} \, G_{M_5}(k_1^2,k_1\cdot k_2,k_2^2)\,,
\label{Gtransition}
\end{align}
where the pseudoscalar meson's momentum $P=k_1+k_2$, $k_1$ and $k_2$ are the photon momenta.  We compute $G_{M_5}$ using the Dyson-Schwinger equations (DSEs) \cite{Cloet:2013jya,  Roberts:2015lja}, a symmetry-preserving framework whose elements have an explicit connection with QCD \cite{Binosi:2014aea}.  At leading-order in this approach (rainbow-ladder, RL, truncation) \cite{Raya:2015gva}:
\begin{align}
\nonumber
T_{\mu\nu}(k_1,k_2) & = \! {\rm tr}\!\!\! \int\!\!\!\frac{d^4 \ell}{(2\pi)^4} \,i {\cal Q} \chi_\mu(\ell,\ell_1)\, \\
&
\times \Gamma_{M_5}(\ell_1,\ell_2) \, S(\ell_2) \, i {\cal Q} \Gamma_\nu(\ell_2,\ell)\,,
\label{anomalytriangle}
\end{align}
where $\ell_{1}=\ell+k_1$, $\ell_{2} = \ell - k_2$, and the kinematic constraints are $k_1^2=Q^2$, $k_2^2=0$, $2\, k_1\cdot k_2=- (m_{M_5}^2+Q^2)$.

In Eq.\,\eqref{anomalytriangle}, $\Gamma_{M_5}$ is the meson's Bethe-Salpeter amplitude \cite{LlewellynSmith:1969az, Krassnigg:2009zh}; ${\cal Q}$ is a matrix that associates an electric charge with each of the meson's valence constituents, whose propagation is described by $S$; and $\Gamma$, $\chi$ are, respectively, amputated and unamputated photon-quark vertices.  The momentum-dependent elements indicated here have long been the subject of careful scrutiny, so that the character of each is deeply understood, and accurate numerical results and interpolations are available.  Many of these things are detailed in Ref.\,\cite{Raya:2015gva}, which examines the $\gamma \gamma^\ast \to \pi^0$ transition and unifies its treatment with that of the pion's elastic electromagnetic form factor \cite{Chang:2013nia}.  We now capitalise on those insights and methods in computing and interpreting the $\gamma \gamma^\ast \to \eta_{c,b}$  transitions.

The dressed $c$- and $b$-quark propagators and $\eta_{c,b}$ Bethe-Salpeter amplitudes in Eq.\,\eqref{anomalytriangle} were computed in Ref.\,\cite{Ding:2015rkn}, and used to predict the leading-twist parton distribution amplitudes (PDAs) of these heavy pseudoscalar mesons and a range of other quantities.  For example, from the decay constants reported therein [$f_{\eta_c}=0.26\,$GeV, $f_{\eta_b}=0.54\,$GeV] one obtains the following widths [$e_{M_5^q}=(2/3),(-1/3)$ for $\eta_{c,b}$, respectively]:
\begin{subequations}
\label{G0fM5}
\begin{align}
& \Gamma[M_5\to \gamma\gamma]
= \tfrac{1}{4} \pi \alpha_{\rm em}^2 m_{M_5}^3 |G_{M_5}(Q^2=0)|^2\\
\label{G0fM5line2}
& = \frac{8\pi\alpha_{\rm em}^2 e_{M_5^q}^4 f_{M_5}^2}{m_{M_5}}
\left\{\begin{array}{l}
 \stackrel{\eta_c}{=} 6.1\,{\rm keV}\\
 \stackrel{\eta_b}{=} 0.52\,{\rm keV}
\end{array}\,,
\right.
\end{align}
\end{subequations}
where the formula in Eq.\,\eqref{G0fM5line2} is drawn from Ref.\,\cite{Lansberg:2008cq}.  The $\eta_c$ width compares favourably with a world average \cite{Agashe:2014kda}: $\Gamma[\eta_c\to \gamma\gamma] =5.1\pm 0.4\,$keV, but the $\eta_b\to \gamma\gamma$ decay has not yet been seen.  In Eqs.\,\eqref{G0fM5}, one has a useful constraint on $|G_{\eta_{c,b}}(Q^2=0)|$.

\begin{table}[t]
\caption{Interpolation coefficients for PTIRs. \emph{Upper panel}: dressed propagators for $c$- and $b$-quarks [Ref.\,\cite{Raya:2015gva}, Eq.\,(A1)] -- the pair $(x,y)$ represents the complex number $x+ i y$; and \emph{lower panel}: Bethe-Salpeter amplitudes of $\eta_{c,b}$ mesons [Ref.\,\cite{Raya:2015gva}, Eq.\,(A2)--(A4)].  (Mass-dimensioned quantities in GeV).
\label{TablePTIR}
}
\begin{center}
%

\begin{tabular*}
{\hsize}
{
c|@{\extracolsep{0ptplus1fil}}
c@{\extracolsep{0ptplus1fil}}
c@{\extracolsep{0ptplus1fil}}
c@{\extracolsep{0ptplus1fil}}
c@{\extracolsep{0ptplus1fil}}
c@{\extracolsep{0ptplus1fil}}}\hline
& $z_1$ & $m_1$  & $z_2$ & $-m_2$ \\
$c$\phantom{00} & $(0.49,1.12)$ & $(1.75,0.26)$ & $(0.028,0)$ & $(2.36,0)$ \\
$b$\phantom{00} & $(0.49,0.97)$ & $(5.06,0.50)$ & $(0,0.0018)$ & $(2.45,1.91)$ \\\hline
\end{tabular*}

\vspace*{2ex}

\begin{tabular*}
{\hsize}
{
l|@{\extracolsep{0ptplus1fil}}
c@{\extracolsep{0ptplus1fil}}
c@{\extracolsep{0ptplus1fil}}
c@{\extracolsep{0ptplus1fil}}
c@{\extracolsep{0ptplus1fil}}
c@{\extracolsep{0ptplus1fil}}
c@{\extracolsep{0ptplus1fil}}
c@{\extracolsep{0ptplus1fil}}
c@{\extracolsep{0ptplus1fil}}}\hline
    & $c^{\rm i}$ & $c^{u}$ & $\phantom{-}\nu^{\rm i}$ & $\nu^{\rm u}$
    & $a$ & $\Lambda^{\rm i}$ & $\Lambda^{\rm u}$\\\hline
$E_{\eta_c}$ & $\phantom{-}0.88\phantom{8}$ & $\phantom{-}0.15\phantom{15}$ & $3$ & 1
    & $2$ & 1.7 & 0.77\\
$F_{\eta_c}$ & $\phantom{-}0.22\phantom{8}$ & $\phantom{-}0.012\phantom{5}$ & $3$ & 1
    & $3/[\Lambda^{\rm i}_F]$ & 1.5 & 0.73 \\
$G_{\eta_c}$ & $-0.018$ & $-0.0015$ & 3 & 1 & $4.4/[\Lambda^{\rm i}_G]^3$ & 1.3 & 0.92 \\\hline
$E_{\eta_b}$ & $\phantom{-}0.77\phantom{8}$ & $\phantom{-}0.38\phantom{15}$ & $7$ & 1
    & $5$ & 2.5 & 1.0\phantom{0}\\
$F_{\eta_b}$ & $\phantom{-}0.037$ & $\phantom{-}0.013\phantom{5}$ & $7$ & 1
    & $20/\Lambda^{\rm i}_{F}$ & 2.5 & 0.82 \\\hline
\end{tabular*}
\end{center}

\vspace*{-4ex}

\end{table}

It is difficult to reliably compute integrals like that in Eq.\,\eqref{anomalytriangle} on the entire domain of experimentally accessible momentum transfers if the propagators, amplitudes and vertices are only known numerically \cite{Maris:2000sk, Maris:2002mz}.  Therefore, following Refs.\,\cite{Raya:2015gva, Chang:2013nia}, we have developed perturbation theory integral representations (PTIRs) of these elements using the gap and Bethe-Salpeter equation solutions computed in Ref.\,\cite{Ding:2015rkn}.  The PTIRs are fully determined by Eqs.\,(A1)--(A4) in Ref.\,\cite{Raya:2015gva} and the lists in Table~\ref{TablePTIR}; and the functions thus defined serve as accurate, algebraic interpolations of the solutions in Ref.\,\cite{Ding:2015rkn} throughout the domains sampled in evaluating the integral in Eq.\,\eqref{anomalytriangle}.  It is noteworthy that in the pseudoscalar channel: \emph{(a)} heavy-heavy mesons are predominantly $S$-wave in character; and \emph{(b)} the $G$-components of meson Bethe-Salpeter amplitudes correspond to $P$-waves in the bound-state rest-frame.  These observations explain the small size of $G_{\eta_c}$, defined by the values of $c^{i,u}$ in Row~3, lower-panel, Table~\ref{TablePTIR}, and the complete neglect of $G_{\eta_b}$.

Only the photon-quark vertices in Eq.\,\eqref{anomalytriangle} remain unspecified.
Since we pursue a unified treatment, we use precisely those forms detailed in Refs.\,\cite{Raya:2015gva, Chang:2013nia}, with the Breit-frame kinetic energies ${\mathpzc E}_{\eta_{c,b}} = (\tfrac{1}{4}Q^2+m_{\eta_{c,b}}^2)^{1/2}-m_{\eta_{c,b}}$ and the momentum redistribution factors fixed via Eq.\,\eqref{G0fM5}: ${\mathpzc s}_0^{c}=0.78$, ${\mathpzc s}_0^{b}=0.23$.

\smallskip

\noindent\emph{III.$\;$Transition Form Factors: Calculation}\,---
%
With all elements in Eq.\,\eqref{anomalytriangle} now expressed via a generalised spectral representation, computation of $G(Q^2)$ reduces to the task of summing a series of terms, all of which involve a single four-momentum integral.  The integrand denominator in every term is a product of $\ell$-quadratic forms, each raised to some power.  Within each such term, one uses a Feynman parametrisation in order to combine the denominators into a single quadratic form, raised to the appropriate power.  A suitably chosen change of variables then enables routine evaluation of the four-momentum integration using algebraic methods.  After calculation of the four-momentum integration, evaluation of the individual term is complete after one computes a finite number of simple integrals; namely, the integrations over Feynman parameters and the spectral integral.  The complete result for $G_{\eta_c}(Q^2)$ follows after summing the series.

\begin{figure}[t]
\centerline{\includegraphics[clip,width=0.42\textwidth]{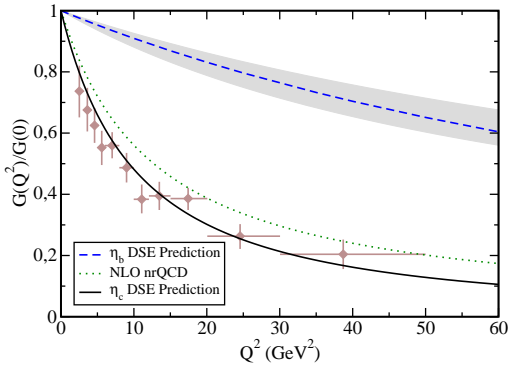}}
\caption{\label{figTranFF}
Transition form factors: $\gamma\gamma^\ast \to \eta_f$, $f=c,b$.
Curves:
    dashed (blue) curve -- our prediction for $\eta_b$;
    (grey) band -- NNLO nrQCD result for $\eta_b$ \cite{Feng:2015uha} (the band width expresses the sensitivity to the factorisation scale);
    dotted (dark-green) curve -- NLO nrQCD result for $\eta_c$ \cite{Feng:2015uha}  (the NNLO result is omitted because it is vastly different from the data and exhibits marked sensitivity to the assumed factorisation and renormalisation scales); and
    solid (black) curve -- our prediction for $\eta_c$.
Data from Ref.\,\cite{Lees:2010deBaBar}.}
\end{figure}

Our prediction for $G_{\eta_c}(Q^2)$ is displayed in Fig.\,\ref{figTranFF}.  It was obtained via the procedure detailed above supplemented by the inclusion of leading-order ERBL (QCD) evolution \cite{Lepage:1979zb, Efremov:1979qk, Lepage:1980fj} of the meson's Bethe-Salpeter amplitude, the nature and necessity of which is described in Ref.\,\cite{Raya:2015gva}.  In this case such evolution produces a $Q^2$-dependent \emph{enhancement}, which grows logarithmically from $1.0$ on $Q^2\simeq 0$ to a value of $\approx 1.05$ at $Q^2=60\,$GeV$^2$, \emph{i.e}.\  on the domain depicted, it has an noticeable and important impact.  We note, too, that $\Gamma[\eta_c\to\gamma\gamma] = 5.1\pm0.4\,$keV can be obtained using ${\mathpzc s}_0^{c}=0.67\pm0.04$, but this value yields a practically identical result for the $Q^2$-dependence of $G_{\eta_c}(Q^2)$: the curves agree within a line-width.  Our result for the $\eta_c$ interaction-radius is $r_{\eta_c} = 0.16\,$fm$\,=0.23 r_{\pi^0}$, computed from the slope of the transition form factor.  Experimentally \cite{Lees:2010deBaBar}, $r_{\eta_c} = 0.17 \pm 0.01\,$fm.

No parameters were varied in order to obtain the solid curve in Fig.\,\ref{figTranFF}.
The evident agreement with the data from Ref.\,\cite{Lees:2010deBaBar} is therefore invested with considerable meaning.
For example, the prediction derives from an $\eta_c$ Bethe-Salpeter amplitude that produces a leading-twist PDA for this meson, $\varphi_{\eta_c}$, which is piecewise convex-concave-convex and much narrower than the conformal limit result $\varphi^{\rm cl}(x)=6 x (1-x)$.   Hence, the favourable comparison with data confirms the associated prediction for $\varphi_{\eta_c}$ in Ref.\,\cite{Ding:2015rkn}.
In addition, the framework used to produce $G_{\eta_c}(Q^2)$ is precisely the same as that employed for the pion transition form factor in Ref.\,\cite{Raya:2015gva}, and, consequently, the two transitions are simultaneously explained.  Hence, agreement herein with the data from Ref.\,\cite{Lees:2010deBaBar} may equally be interpreted as confirmation of the results in Ref.\,\cite{Raya:2015gva}, from which it follows that the $\gamma\gamma^\ast \to \pi^0$ data in Ref.\,\cite{Uehara:2012ag} should be considered as the most reliable available measurement of this transition on $Q^2\gtrsim 10\,$GeV$^2$.  It follows that the agreement in Fig.\,\ref{figTranFF} between data and our result provides further support for the prediction in Refs.\,\cite{Chang:2013pq, Cloet:2013tta, Horn:2016rip}, \emph{viz}.\ at scales typical of modern hadron physics, the pion's leading-twist PDA is dilated, such that, unlike $\varphi_{\eta_c}$, $\varphi_{\pi}$ is significantly broader than $\varphi^{cl}$.
Furthermore, the agreement between data and our prediction, the qualitative agreement between both and the NLO nrQCD result, and the disagreement between data and our result on one hand, and the NNLO nrQCD result on the other, suggest that one must seriously question the usefulness of nrQCD in applications to exclusive processes involving charmonia \cite{Feng:2015uha}.

It is natural at this point to consider the asymptotic behaviour of the $\pi$ and $\eta_c$ transition form factors, which can be determined following Ref.\,\cite{Lepage:1980fj}, \emph{viz}.\ $\forall Q^2 \gg 1\,$GeV$^2$
%
\begin{equation}
\label{Q2clAll}
Q^2 G_{M_5}(Q^2) = \mathpzc{u}_{M_5} f_{M_5}\int_0^1 dx\, \varphi_{M_5}(x;Q^2)/x\,,
\end{equation}
where $M_5 = \pi, \eta_c$ and $\mathpzc{u}_{\pi}=2$ \emph{cf}.\ $\mathpzc{u}_{\eta_c}=8/3$ reflects differences between the electric charges of the relevant valence-quarks.  Since $\varphi_{\pi,\eta_c}\to \varphi^{\rm cl}$ in the conformal limit,
\begin{equation}
\label{clRQ2}
\lim_{Q^2\to\infty} [ R(Q^2):=G_{\eta_c}(Q^2) / G_\pi(Q^2) ] = \frac{4}{3} \frac{f_{\eta_c}}{f_\pi}\,.
\end{equation}

\begin{figure}[t]
\centerline{\includegraphics[clip,width=0.42\textwidth]{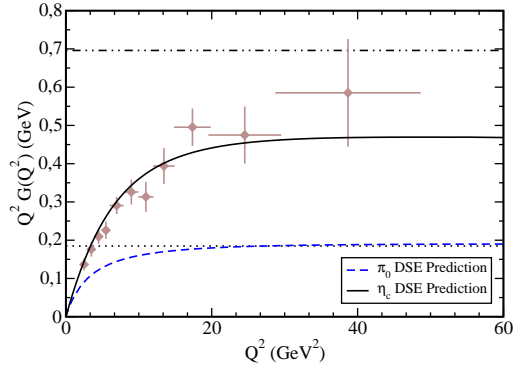}}
\caption{\label{figCL}
$Q^2 G_{M_5}(Q^2)$ for $\gamma\gamma^\ast \to M_5$, $M_5=\pi^0,\eta_c$.  Dotted and dot-dot-dashed curves display the respective conformal limits, Eq.\,\eqref{Q2clAll}.
$\eta_c$ data drawn from Ref.\,\cite{Lees:2010deBaBar}.  $\pi^0$ data are omitted owing to controversy at large-$Q^2$ \cite{Roberts:2010rn, Brodsky:2011yv, Balakireva:2011wp, Stefanis:2012yw, Bakulev:2012nh, ElBennich:2012ij, Raya:2015gva}.}
\end{figure}

In our unified treatment, we can evaluate this ratio:
$R(Q_{35}^2 := 35\,$GeV$^2$)=2.4.
Extant data are consistent with our prediction, which, however, is just $64$\% of the conformal value in Eq.\,\eqref{clRQ2}.
This mismatch occurs despite the fact that $Q_{35}^2 G_\pi(Q_{35}^2) \approx 2f_\pi$.  The discrepancy thus owes to $G_{\eta_c}(Q^2)$.  Digging deeper, one finds that with $\varphi_{\eta_c}(x;Q^2)$ being much narrower than $\varphi^{\rm cl}(x)$, a representation of $\varphi_{\eta_c}(x;Q^2)$ in terms of eigenfunctions of the one-loop ERBL operator must involve a large, negative first ``subleading'' coefficient.  It is then unsurprising that $G_{\eta_c}(Q^2)$ should lie far below its conformal limit value at currently accessed momentum transfers.  (For the pion, the same correction is smaller in magnitude and positive.)  Moreover, given that ERBL evolution is logarithmic, this must remain the case even at $Q^2 \gtrsim1000\,$GeV$^2$ \cite{Cloet:2013jya}.
(With this prediction we contradict the sum rules study in Refs.\,\cite{Lucha:2012ar}, which employs adjustable parameters.)
These points are illustrated in Fig.\,\ref{figCL}.  It is worth remarking that the mismatch between our computed $\eta_b$ transition form factor and its conformal limit ($2f_{\eta_b}/3$) is even more noticeable on this $Q^2$-domain.

A physical context is readily established for these predictions.  Since $m_\pi^2/Q^2_{35} = 0.0008$, $m_\rho^2/Q_{35}^2=0.02$, the hard-photon ``perceives" an almost scale-free system and the $\pi^0$ transition form factor lies in the neighbourhood of its conformal limit.  On the other hand, $m_{\eta_c}^2/Q^2_{35} = 0.4$, $m_{\eta_b}^2/Q^2_{35} = 3.5$, values which reveal that mass-scales intrinsic to the related transitions are commensurate with ($\eta_c$) or greatly exceed ($\eta_b$) those of the probe.  Consequently, at accessible momenta, $\gamma\gamma^\ast\to \eta_{c,b}$ cannot possibly match expectations based on conformal symmetry.

Our prediction for $G_{\eta_b}(Q^2)$ also appears in Fig.\,\ref{figTranFF}.  In this case the computational procedure is indirect because the $b$-quark-related PTIRs defined by Table\,\ref{TablePTIR} are inadequate to the task of eliminating all spurious singularities from the vast integration domain explored by the integral in Eq.\,\eqref{anomalytriangle} when $m_{M_5}^2=m_{\eta_b}^2= 88\,$GeV$^2$ \cite{Agashe:2014kda}.  They are nevertheless quite efficient, allowing a direct computation of the transition form factor on $m_{M_5}^2 \leq 71\,$GeV$^2$.  We therefore computed a pseudo-$\eta_b$ transition form factor as a function of $m_{M_5}=m_{\eta_b^p} \in [7.0,8.4]\,$GeV.  Then, at each value of $Q^2$, the on-shell form factor was determined by extrapolation of the $\eta_b^p$ results, treated as a function of $m_{\eta_b^p}$.  Pad\'e-approximants of order $[k,k]$, $k=1,2,3,4$, were employed, with the difference between their extrapolated values being used to estimate the error in the procedure.  That error is small, lying within the line-width of the dashed (blue) curve in Fig.\,\ref{figTranFF}.
The interaction radius is $r_{\eta_b}=0.041\,$fm$\,= 0.26\, r_{\eta_c}$.  Notably, the ordering of radii follows the pattern:
$r_{\eta_c}/r_{\pi^0} \approx M^E_u/M^E_c$,
$r_{\eta_b}/r_{\eta_c} \approx M^E_c/M^E_b$,
where $M^E_q$ is the Euclidean constituent-quark mass \cite{Ding:2015rkn}, a quantity similar to the $\overline{\rm MS}$-mass often used in connection with heavy quarks.

Concerning $\gamma\gamma^\ast\to \eta_b$, it is worth remarking that the differences between LO, NLO and NNLO nrQCD results are modest \cite{Feng:2015uha}, suggesting that this effective field theory might be a useful tool in connection with analyses of exclusive processes involving bottomonia.  That possibility is supported by the fact that our prediction for this transition form factor lies completely within the (grey) band demarcating the NNLO nrQCD result.

\smallskip

\noindent\emph{IV.$\;$Conclusion}\,---
This study of $\gamma \gamma^\ast \to \eta_{c,b}$ transitions achieves, within a single computational framework that possesses a traceable connection to quantum chromodynamics (QCD), the unification of a wide variety of ground-state $^1S_0$ quarkonia properties -- masses, decay constants, parton distribution amplitudes (PDAs), transition form factors, etc.\ -- with an even wider array of properties of QCD's archetypal Goldstone modes, extending to, \emph{e.g}.\ $\pi\pi$ scattering lengths \cite{Bicudo:2001jq, Bicudo:2001jq}, and electromagnetic pion elastic and transition form factors \cite{Chang:2013nia, Raya:2015gva}.  No parameters were varied in order to achieve agreement with the experimental value of any quantity discussed herein and hence the computed results may validly be described as predictions.
The calculations are built upon the leading-order term in a systematic, symmetry-preserving truncation of those equations in quantum field theory which describe bound-states, their constituents, and the interactions of those constituents with electromagnetic probes.  Quantitative corrections to the results must therefore be expected; but in the channels upon which this study focuses, those corrections are known to be small for reasons that are well understood \cite{Roberts:2015lja}.

The predicted $\gamma \gamma^\ast \to \eta_c$ form factor, $G_{\eta_c}(Q^2)$, matches available data. 
It is thus significant that on $Q^2\lesssim 100\,$GeV$^2$,  $Q^2 G_{\eta_c}(Q^2)$ does not exceed 70\% of the conformal limit result.
%
We attribute this behaviour to compression of the leading-twist PDAs describing heavy-heavy bound-states \emph{cf}.\ the conformal limit.  Such compression is anticipated \cite{Ding:2015rkn}, but the agreement between our predictions and data provides quantitative confirmation.

In confirming the data \cite{Lees:2010deBaBar} as a reliable measure of the $\gamma \gamma^\ast \to \eta_{c}$ transition form factor, our study strengthens claims \cite{Feng:2015uha} that non-relativistic QCD (nrQCD) is not a reliable effective field theory for analyses of exclusive processes involving charmonia.  Regarding $\gamma \gamma^\ast \to \eta_{b}$, on the other hand, there is good agreement between our prediction and the result obtained at next-to-next-to-leading order in nrQCD.  Thus, in this case one should view the theoretical predictions as well-founded and look for them to be verified at a new generation of $e^-\,e^+$ colliders \cite{Nature:2014China}.

One can now draw various threads together and argue that, with the predictions described herein, we have reached a sound understanding of the distribution of valence-quarks within mesons, a picture which smoothly joins Goldstone modes, constituted from the lightest-quarks in Nature, with systems containing the heaviest valence-quarks that can today be studied experimentally.  Data confirms both that the PDAs of light-quark mesons are dilated with respect to the conformal limit and those for heavy-heavy systems are compressed, becoming narrower as the current-mass of the valence-quarks increases.  (In this connection, the boundary between light and heavy lies just above the strange-quark mass \cite{Ding:2015rkn}.)  These visible features express basic facts about the origin and manifestation of mass within the Standard Model \cite{Roberts:2016vyn}.  Namely, in systems formed by those quarks with the weakest coupling to the Higgs boson, dynamical mass generation via strong-interaction processes (dynamical chiral symmetry breaking -- DCSB) is the dominant effect, and it is revealed in a marked dilation of the PDAs associated with these systems \cite{Chang:2013pq}.  Moreover, DCSB ensures that this dilation persists even when coupling to the Higgs vanishes.  On the flip-side, the leading-twist PDA for a system constituted from valence-quarks with a strong-coupling to the Higgs is narrow, it becomes narrower as that coupling increases, and there is no mass-scale within the Standard Model which can prevent the PDA approaching a $\delta$-function as the Higgs coupling continues to grow.   It follows that the root-mean-square relative
velocity of valence-constituents within a meson has a nonzero, finite upper bound, fixed by the strength of DCSB, but must vanish with increasing current-mass of the meson's valence-quarks.

\smallskip

\noindent\textbf{Acknowledgments}.
%
This research was enabled and facilitated by the
\emph{2$^{\it nd}$ Sino-Americas Workshop and School on the Bound-State Problem in Continuum QCD}, Central China Normal University, Wuhan, China, 16-20 November 2015.
Research supported by:
CIC (UMSNH) and CONACyT Grant nos.\ 4.10 and CB-2014-22117;
and
U.S.\ Department of Energy, Office of Science, Office of Nuclear Physics, under contract no.~DE-AC02-06CH11357.


\end{document}